\documentstyle[11pt,aaspp4,psfig]{article}  
 
\begin{document}

\title{The shortest-period M-dwarf eclipsing system BW3~V38}

\author{\sc Carla Maceroni\altaffilmark{1}\\
\rm Electronic-mail: {\it maceroni@coma.mporzio.astro.it\/}}

\affil{Rome Observatory, 
via dell' Osservatorio 2, I-00040 Monteporzio C., Italy
}

\and

\author{\sc Slavek M. Rucinski\altaffilmark{2}\\
\rm Electronic-mail: {\it rucinski@astro.utoronto.ca\/}}
 
\affil{81 Longbow Square, Scarborough, Ontario M1W~2W6, Canada}
\altaffiltext{1}{Affiliated with the Institut
d'astrophysique de Paris -- CNRS, France}
\altaffiltext{2}{Affiliated with the Department of Astronomy,
University of Toronto and Department of Physics and Astronomy,
York University}

\centerline{\today}

\begin{abstract}
The photometric data for a short-period (0.1984 day)
eclipsing binary V38 discovered by the OGLE 
micro-lensing team in Baade's Window field BW3 have been analyzed.
The de-reddened color $(V-I_C)_0=2.3$ 
and the light-curve synthesis solution of 
the $I$-filter light curve suggest a pair of strongly-distorted
M-dwarfs, with parameters between those of YY~Gem and CM~Dra,
revolving on a tightest known orbit among binaries consisting of
 Main Sequence stars. The primary, more massive and hotter, component
maybe filling its Roche lobe.
The very small amount of angular momentum in the orbital motion
makes the system particularly important for studies of angular
momentum loss at the faint end of the Main Sequence. Spectroscopic
observations of the orbital radial velocity variations as well as of
activity indicators are urgently needed for a better
understanding of the angular-momentum and internal-structure
evolutionary state of the system.
\end{abstract}
 

\section{INTRODUCTION}
\label{into}

The authors of this paper share a strong conviction that light-curve
solutions of individual eclipsing binary stars should belong to the
past, with the exception of extraordinarily important systems. We feel
that the variable star V38 in the third (BW3) field of the 
OGLE micro-lensing
project is such a system. With the orbital period of only 0.1984 day, 
it is the shortest-period known binary consisting of Main Sequence
stars.

BW3~V38, from now on called \#3.038 ($\alpha_{2000} = 
18^h04^m44\fs 19$, $\delta_{2000} = -30\arcdeg 09\arcmin 05\farcs1$, 
$P = 0.19839$ day, $I_{max} = 15.83$,
$(V-I_C)_{max}=2.45$, $\Delta I = 0.78$),  
appears in the second instalment of the OGLE Catalog of periodically
variable stars (Udalski et al.\markcite{uda2} 1995) among contact
binaries of the W~UMa-type (EW). However, it did not pass the Fourier
filter and was not included in the impersonally-selected sample of
contact binaries in the discussion of all contact systems in Baade's
Window (the R-sample) by Rucinski\markcite{ruc97} (1997 = R97). 
But this rejection was a marginal one because the Fourier filter had been 
constructed for the most common G-type systems observed in
the $V$-band, whereas in this case, 
it was applied to a M-type  system observed in
the $I$-band. It can be argued that light
curves of contact binary systems
are primarily dependent on the strongly perturbed geometry of
the stars, and weakly dependent on the atmospheric properties
(Rucinski\markcite{ruc93} 1993), but it is possible that application of
the Fourier filter was carried too far for the case of \#3.038. A full
light-curve solution seemed to be in order.

The importance of \#3.038 stems from its very short period, but also,
if it is a contact system, from its very late spectral
type. Short-period, late-type systems are common among contact
binaries, but the period distribution of these systems
shows a sharp cutoff, which is
well defined by the system CC~Comae, with the orbital period is 0.221 day
and the intrinsic color $V-I_C = 1.39$ 
(Bradstreet \markcite{br85} 1985).
 In the volume-limited sample to 3 kpc formed from the OGLE
sample (R97), contact binaries appear with high apparent frequency
only within intervals of orbital periods and intrinsic colors 
$0.25 < P < 0.5$ day and $0.4 < V-I < 1.2$. This color range translates
into the range of spectral types of about F2 to K5.
It is an interesting puzzle of the stellar
structure theory why the short period cutoff is so well defined and sharp,
and why there are no contact binaries
consisting of late-K and M dwarfs. Some arguments
have been presented in favor of the stellar structures converging to
the fully-convective state (Rucinski\markcite{ruc92}
1992), but -- quite possibly --
the reasons for non-existence of very low-mass systems are different
and more complex. The discovery of \#3.038 possibly confronts us 
with a contact system beyond the current period cutoff. The
spectral type of \#3.038 must be much later than that of 
CC~Com, as the observed colors of the two systems 
are $V-I_C = 2.45$ and 1.39,
respectively. A simple estimate of the reddening based on the
maps of the reddening of Stanek\markcite{sta96} (1996) (see
Sec.~\ref{assum})
suggests $E_{V-I} = 0.18$. Thus, \#3.038 is much cooler than CC~Comae
and consists of M-type dwarfs.

Even if \#3.038 is not a contact system, it is extremely interesting
object as a very close binary consisting of M-type dwarfs. 
Most stars in the Galaxy are M-type dwarfs, yet --
because of their faintness -- we know very little about them. In fact,
the low end of the Main Sequence is calibrated by only two eclipsing
systems, YY~Gem and CM~Dra. Because of large numbers of the M-type
dwarfs, small systematic errors made in analyses of these two stars
may have profound effects in our understanding of even so remotely related
subjects as dynamics of the Galaxy or the critical density of
the Universe. In addition, analysis of old, late-type dwarfs can
give us such basic data as the primordial helium and metal abundances
(Paczy\'nski \& Sienkiewicz \markcite{pacz84} 1984; 
Chabrier \& Baraffe \markcite{ChB} 1995; 
Metcalfe et al. \markcite{met} 1996). 
We should note that CM~Dra is a Population~II object while YY~Gem
belongs to a sextuple $\alpha$~Gemini (Castor) system which must be
relatively young (age about 500 Myr)
as it contains an unevolved A1V star. Thus, 
we urgently need more systems to clearly understand effects of age and 
metallicity in the lower Main Sequence.

The strong distortion of
components is directly visible in the light curve 
(Figure~\ref{fig1}) of \#3.038. 
With continuing angular-momentum loss, which must operate in such 
a late-type system, the binary
is apparently on its way to become a contact system. The major
question is: Why we do not see such systems in large numbers? 
Is the angular momentum loss so severe that contact systems very
quickly merge and form single stars? Or, at the other extreme, is the
loss so feeble that most among close M-dwarf pairs did not lose enough
angular momentum in the lifetime of the Galaxy to form contact
systems? If the latter is the case, is 
\#3.038 one of the oldest among such close M-type binaries? 

This paper is meant as a first, exploratory attempt to extract as
much information from the extant data as possible. We hope to
stimulate interest in this important system; in particular we hope
that the much needed spectroscopic observations will be obtained soon. 

\section{THE LIGHT CURVE SOLUTION}

\subsection{Assumptions}
\label{assum}

\begin{figure}      
\centerline{\psfig{figure=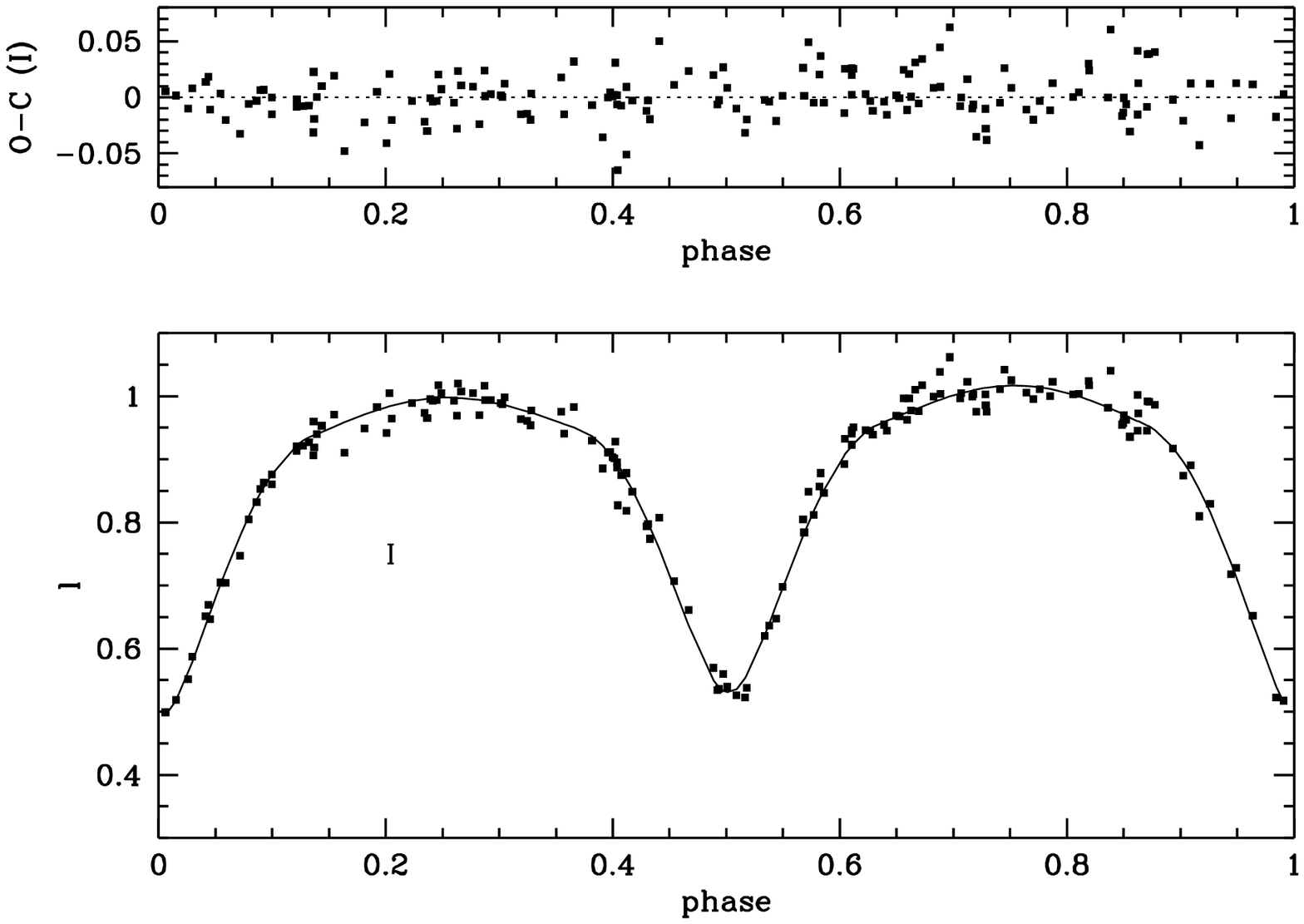,height=6.0in}}
\vskip 0.5in
\caption{\label{fig1}
Light curve of \#3.038 in the $I$-band from the OGLE database. The
continuous line gives the best fit. The deviations are shown in the
panel above the light curve.}
\end{figure}

While attempting a more detailed study of \#3.038, one is confronted
with the limited extent of the 
available data. The OGLE data published in the catalogs and available
over the {\it ftp\/} service consist of the light curve in the
$I$-band, the maximum-light $V-I_C$ color and the orbital period.
The light curve (Figure~\ref{fig1}) is rather noisy, 
with a standard error
at maxima of $\simeq 0^m\!.02$, but with deep eclipses 
($\Delta I = 0.78$) suggesting a high value of the inclination.
An  analysis of the light curve is still sensible, but one has to be
careful about the strategy to apply in the solution
and about the estimation of uncertainties
of the derived parameters. With any standard light-curve synthesis
method, such as the Wilson and Devinney\markcite{WD71} (1971 = WD)
program (we used its
updated 1992 version), the main and the most obvious concern would be
the choice and the number of free parameters. We address this matter
below, but first we look into the available information about
components of \#3.038.

\placefigure{fig1}

The observed, possibly strongly-reddened $V-I_C$ color is 
the only information about the spectral type of the components. To
derive the unreddened $V-I_C$ color and the approximate
absolute magnitude of
the system, we used a simple iterative procedure which
followed the one presented in R97. The R97 computations
were based on  Stanek\markcite{sta96} (1996) 
reddening maps of Baade's window
and on the calibration period-color-magnitude 
established for W~UMa binaries.
We do not  know whether  \#3.038 is actually a contact system.
Therefore, we used the same iterative approach as in R97, 
but  replaced the W~UMa calibration with
the Main Sequence $M_V=M_V(V-I_C)$ relation
 of Reid and Majewski \markcite{rm93} (1993),
shifted upward by 0.75 mag for a binary system of identical components.
The  final values are: $(V-I_C)_0=2.3 \pm 0.1$, assuming errors of
0.03 and 0.5 in $V-I_C$ and $M_I$, respectively.
The predicted systemic absolute magnitude is $M_I \simeq 7.5$ and the 
distance is $d= 400 \pm 85$ pc. The relatively red intrinsic
color corresponds to a M3 spectral type,  so that
a reasonable value of primary effective temperature  is 
$T_1= 3500$ K (Bessell 1979 \markcite{Be1}, 1990\markcite{Be2}).

Extensive tables of limb darkening coefficients for R, I, J, K bands have
been recently published by Claret et al.\ \markcite{cl95}(1995). We
used the (linear) limb darkening coefficient, $x=0.56$, corresponding to the
primary effective temperature and $\log g$ between 4 and 5.   
The gravity darkening and reflection coefficient  were fixed at the
standard values for convective envelopes ($g=0.32$, $A=0.5$;
Lucy \markcite{lu67}1967, Rucinski \markcite{ruc69}1969).
The low effective temperature of primary does not allow the use of the
model-atmosphere option in the WD 
code which is limited to the temperature range 
4,000 -- 25,000~K (we note that this option in the 1992 version of the
code uses the obsolete model atmospheres from
1969). All computations were therefore performed 
in the black body approximation. For a one band-pass light-curve in
$I$, this is definitely not a critical assumption.

The maxima of the light curve show slightly different heights. The
difference is about $\delta I = 0.02$ which can be modeled by a dark
spot on  the larger star. In order not to introduce too many free
parameters into the geometric-elements solution, the 
spot parameters were determined 
by trial an error  {\it before} starting the differential correction 
procedure. Since the spot characteristics are 
very weakly constrained by light curve solutions without 
spectroscopic information (Maceroni and Van't Veer \markcite{MV93}1993),
we assumed the spot to be located on the equator of the primary 
component. The other fixed
parameters were: longitude and angular radius of the spot 290
and 11 degrees, respectively, and the temperature factor 
$T_{\rm spot}/T_{\rm star}=0.8$.
The advantage of having a spot is in preventing 
occurrence of oscillations in the iterative procedure between solutions 
which fit one maximum at a time.

\subsection{Results of the solution}

\begin{figure}      
\centerline{\psfig{figure=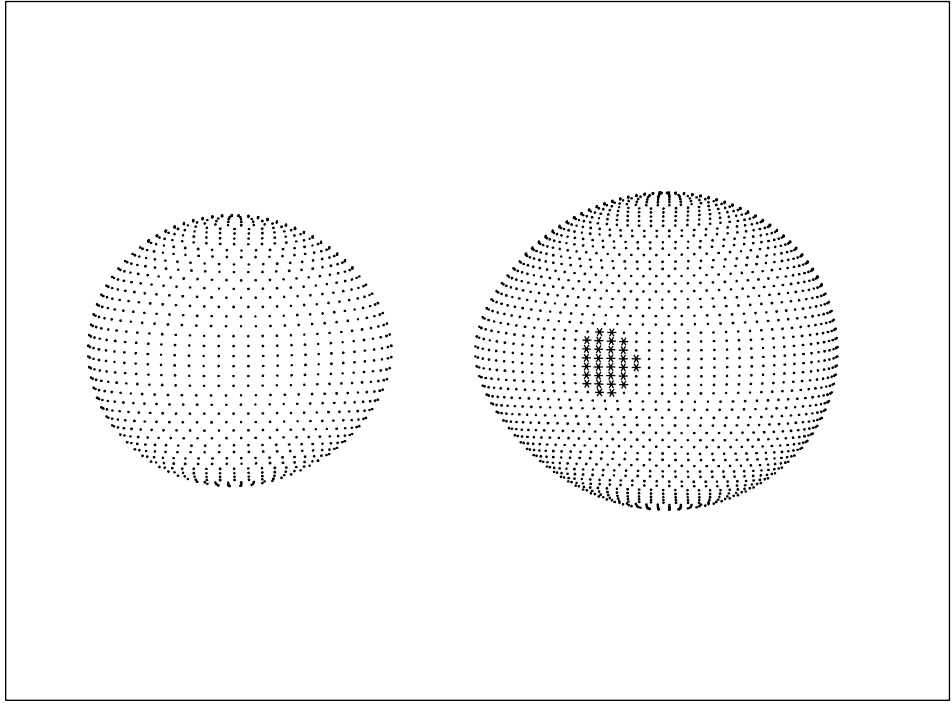,height=6.0in}}
\vskip 0.5in
\caption{\label{fig2}
The three-dimensional picture of \#3.038 at the orbital phase 0.25.}
\end{figure}

The input model for the light curve solution consisted of a system of two
identical stars with the characteristics described 
in the previous section  and exactly 
filling the inner critical Roche surfaces. From this point, the
 surface potentials were left  
the freedom to vary in an independent way, adopting the
WD ``mode 2'' of operation, which is normal for detached systems.
We used seven adjustable parameters: the orbital inclination $i$ (in
degrees), the mass ratio of the system $q$ (expressed as 
the mass of the eclipsing
component to the eclipsed one in the deeper minimum), the two surface
equipotentials $\Omega_1$ and $\Omega_2$ (defined as in the WD code),
the secondary effective temperature $T_2$, the primary component
luminosity $L_1$ in the spectral band used and the phase shift of
the primary eclipse $\Delta \phi$ (expressed in units of
the orbital period).
Their subdivision in two subsets (Wilson and Biermann
\markcite{WB76}1976) helped to control
the strong inter-dependencies 
present among some of the parameters during the iteration process.
The procedure quickly converged   to the solution presented in
Table~\ref{tab1}. This table contains the main solution as well as the
results of an ``bootstrap re-sampling'' experiment described below.
The geometry of the final model is presented in Figure~\ref{fig2}. 
The two components are very close to, but still inside, their
respective Roche critical surfaces. However, for the more massive and
hotter component, the results are consistent at the one-sigma level,
with the star actually filling its critical inner potential, popularly
known as the ``Roche lobe''.

\placetable{tab1}

\placefigure{fig2}

\subsection{Uncertainties of parameters}
\label{uncert}

It is well known in the community of users of the Wilson-Devinney
program that this fine code provides excellent light curves, but that
the estimates of errors of the adjustable parameters are 
unrealistically small. The reason is partly the strong correlation
between the relatively many parameters, partly the non-normal
distribution of measurement errors. The WD code provides the ``probable''
errors, $\epsilon_i$, which are derived by the differential correction
routine and are related to the standard errors of the linearized
least-squares algorithm through $\epsilon_i=0.6745 \sigma_i$.
The errors $\sigma_i$ can be used as a measure of the 
uncertainties only for normal distributions of the errors.

As explained in the book ``Numerical Recipes'', Sec.15.6 (Press et al.\
\markcite{nr92} 1992), 
the technique of bootstrap re-sampling is probably the most useful
for estimation of confidence levels for complex least-squares
solutions. It uses input data re-sampled with repetitions, establishing
confidence levels from the spread in resulting parameters.
However, its application was not possible with the
current WD code, which -- by intent -- was
designed on the principle of ``interactive branching'' during the
differential correction procedure (Wilson
\markcite{Wil85}1985). Combining the interactive branching with the
bootstrap re-sampling would 
mean performing thousands of separate solutions ``by
hand''. Instead, we used a
 simplified bootstrap approach in which 10,000 least-squares
solutions were made within the minimum already established by a
single, iterated solution. Thus, only one set of light-curve
deviations and parameter derivatives were used. This, by necessity,
would give us under-estimates of the
real errors. However, as we 
have found through application of this approach, 
these estimates are larger that the formal mean standard
errors and do show strong inter-parametric correlations. We should
remember that inter-parametric correlations are not always a
detrimental circumstance, as some derived parameters (such as radii)
may show smaller uncertainties than for the case of simple quadrature
addition of uncorrelated errors. The bootstrap-estimated errors tend
to be also somewhat pessimistic as some re-samplings give light
curves so clumpy and unevenly covered that nobody would ever attempt
solving them. Since we see potential for under- and over-estimation of
the errors, we give in Table~\ref{tab1} 
both sets of error estimates, the mean standard errors for the
iterated solution and the one-sigma ranges around the
median values for the bootstrap experiment.

\begin{figure}      
\centerline{\psfig{figure=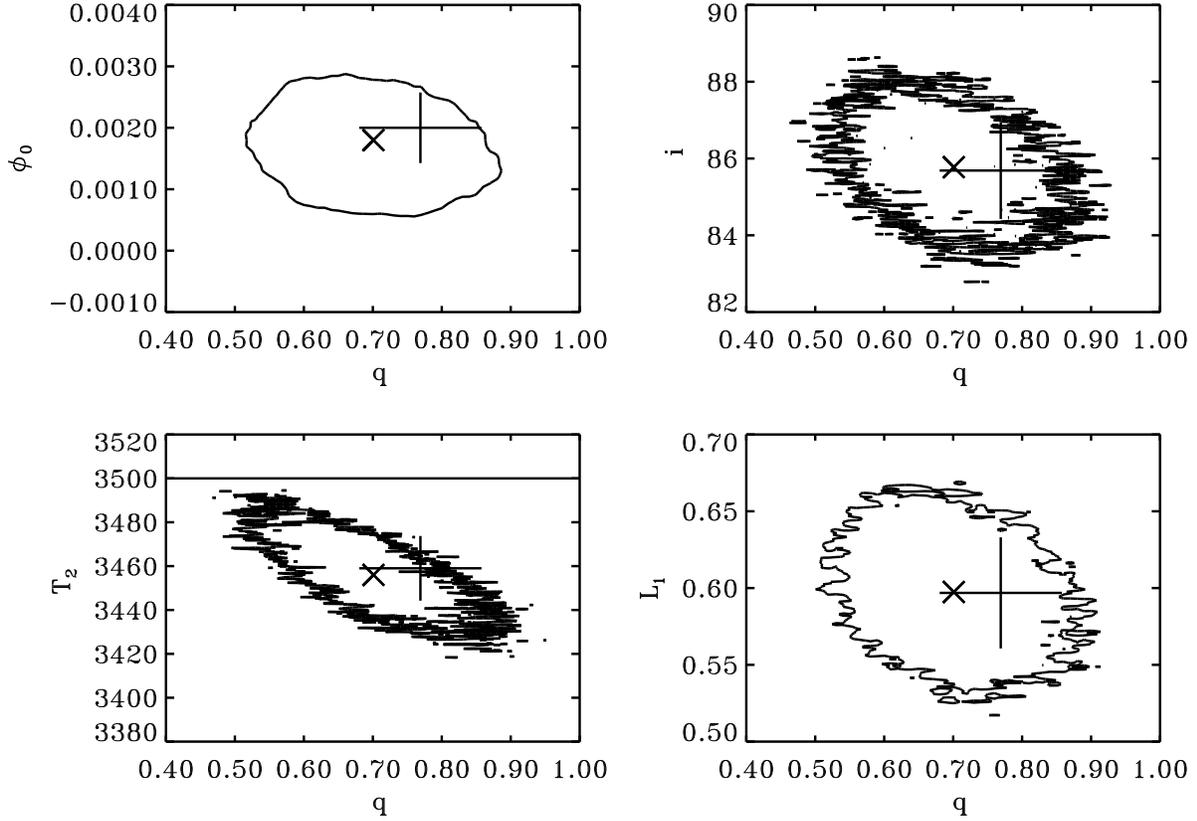,height=4.5in,angle=90}}
\vskip 0.5in
\caption{\label{fig3}
Results of 10,000 ``bootstrap re-sampling'' solutions, as described in
the text. The contours define the one-sigma uncertainty levels
by enclosing 68.3\% of all cases. The poor definition
of some of the contours is entirely due to the contouring routine.
The large crosses give the results of the mean iterated solution, with
lengths of arms equal to the standard errors of parameters. The small
slanted crosses give the median values for individual
parameters. Refer to the text for the reasons why the mean and median
values do not coincide. Note that the temperature of the secondary
component was obtained by assumption of the fixed temperature of the
primary, $T_1 = 3500$ K.}
\end{figure}

A selection of contour plots giving two-dimension representations of
the one-sigma uncertainty levels is shown in Figures~\ref{fig3} and
\ref{fig4}. They have been obtained by finding contours enclosing
 68.3\% of the bootstrap solutions around their respective median
values. They are frequently elliptical
(indicating inter-parametric correlations), but symmetric around
the medians of the respective
seven primary parameters. The symmetry is due to use of one set of
derivative  within the 
minimum of the summed squared deviations. It should be noted
that the bootstrap solutions tend to cluster slightly away
from the best, iterated solution. 
This is caused by the fact that the bootstrap
solutions are always based on light curves with 
poorer coverage than the full solution. This biases the distribution
of the mass-ratio $q$, as this  essential parameter is determined solely
from the distortion of the light curve between minima. If this part of
the light curve is poorly covered, or if observations have large
errors, then the mass-ratio would tend to take smaller values.
The only
proper resolution of this difficulty would be to obtain spectroscopic
radial-velocity data which would stabilize the solution.

\placefigure{fig3}

\subsection{Relation to the Roche geometry}

\begin{figure}      
\centerline{\psfig{figure=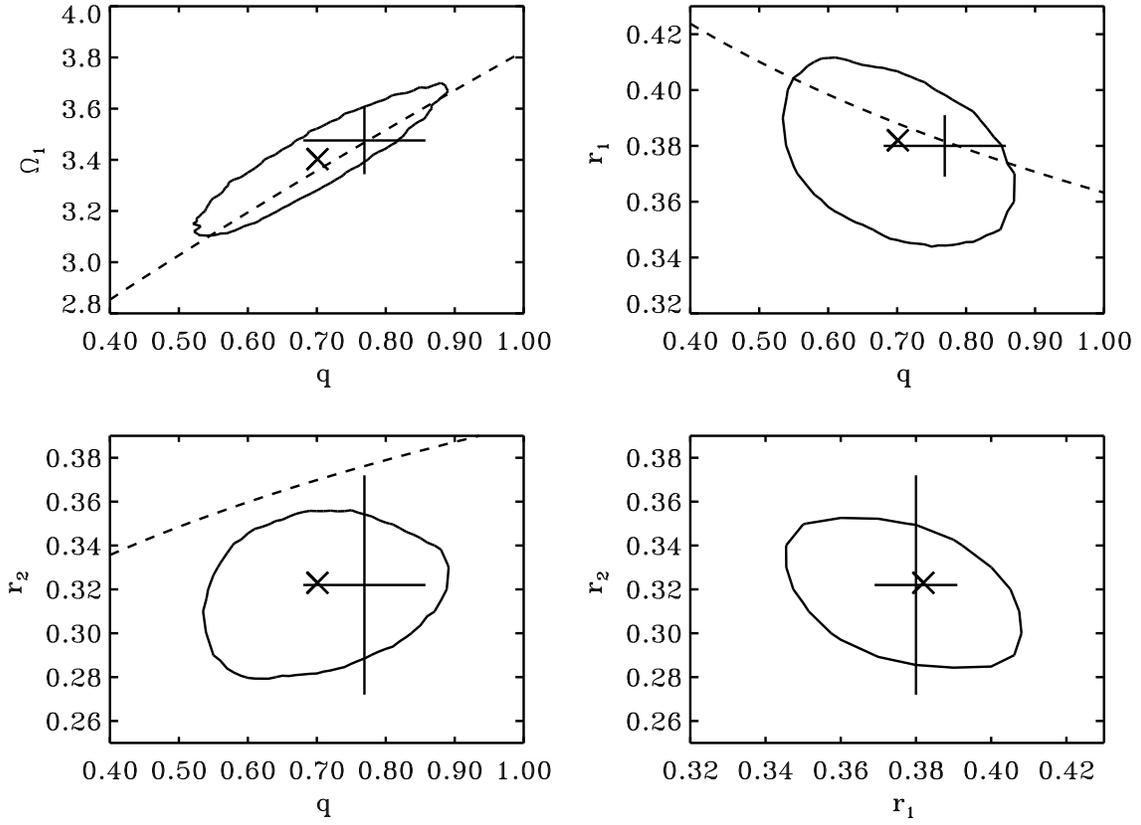,height=4.5in,angle=90}}
\vskip 0.5in
\caption{\label{fig4}
Additional combinations of parameters in the ``bootstrap re-sampling''
solutions. The broken lines refer to  critical (Roche) surface values. 
Since the radii are the derived parameters, their median
values are sometimes shifted from the centers of the one-sigma
contours.}
\end{figure}

The Figure~\ref{fig4} contains uncertainty plots for the relative
radii of the components. Here we used the ``side'' radii in the
orbital plane which -- as can be shown -- are practically equal to the
volume radii. In addition to the mean radii from the iterated
solution, we also show the median values. Note that they are not
exactly in the centers of the one-sigma
contours, as for the other, primary elements. This is due to the
inter-parametric correlations, since the radii are given by
expressions of the type $r_i = r_i (\Omega_i,\, q)$.

\placefigure{fig4}

As we can see in Figure~\ref{fig4}, 
the results for the radii are not well constrained. 
However, a plot in the same figure of the uncertainty contours for 
the surface potential of the larger component, $\Omega_1$, 
indicates that the primary almost fills 
its Roche lobe, as the potential is slightly larger than the
critical one. 
The one-sigma contour follows the critical potential curve for
the relatively large range of mass-ratios permitted by the current
solution. If the primary actually fills its lobe, any possible
mass-transfer between components would be undetectable in the 
current, single-epoch, photometric data. 

\section{DISCUSSION OF \#3.038 PROPERTIES}

\subsection{Comparison to YY~Gem and CM~Dra}

With the photometric solution of the light curve given in
Table~\ref{tab1} we still have rather
modest information about the system. Four parameters result from the
solution ($i$, $q$ and $\Omega_1$, $\Omega_2$ or $r_1$, $r_2$),
two describe the relative luminosity characteristics of the
components ($T_2$ and $L_1$) and two more come from the original
observations ($V-I_C$, $P$). On the basis of these, we can attempt to
place \#3.038 relative to other eclipsing M-dwarfs on the Main
Sequence. There are only two such systems: YY~Gem (Castor~C),
analyzed by Bopp (1974) \markcite{Bo} and by Leung \& Schneider (1978)
\markcite{LS} and  CM~Dra analyzed by 
Lacy (1977) \markcite{La} and by Metcalfe et al.\markcite{Met}(1996).
Both are well detached systems with orbital periods 0.81 day and 1.27
day, and both show only moderate proximity effects, in that no mutual
illumination of the components is visible and
the stars are practically undistorted by tidal forces. 
In this respect, \#3.038 is very different as
this is the first known M-type binary showing strong distortion of
components, directly linked to the very short orbital period of 0.198
day. The ages of the reference stars are very different:
YY~Gem belongs to the Castor ($\alpha$~Gem) system whose 
age must be young as it contains hot, unevolved 
stars\footnote{According to probably the best extant spectral
classification of Garrison \& Beattie \markcite{gb97}(1997), 
the hot components of the
systems have spectral types A1m, A2Va, A2m, A5V:. A simple comparison
with the Pleiades (where the earliest spectral types are B6IV and B8V) 
and Hyades (where the earliest spectral types are A2m and A5)
 with the respective ages of about 80 and 600 Myr, indicates that
the age of sextuple system of Castor must be roughly 500 Myr.},
whereas CM~Dra is a Population~II object. Nothing is known at this 
moment about population characteristics of \#3.038. 

The color of \#3.038, $(V-I_C)_0=2.3 \pm 0.1$, gives us
information about its effective temperature relative to YY~Gem and
CM~Dra. Both reference systems were discussed by Monet et al. (1992)
\markcite{Mo}
in relation to other intrinsically faint stars. In particular, their
Kron--Cousins colors were plotted in Figure~11 of Monet et al.\
which relates the colors to the
effective temperature. These unpublished data had come from 
observations at the U.S. Naval
Observatory\footnote{We are indebted to Dr.\ Conard Dahn for sending
us the data.}: YY~Gem, $V-I_C = 1.92$; CM~Dra, $V-I_C = 2.94$.
Bessell (1990) \markcite{Be2} previously 
measured CM~Dra in the same system and
obtained $V-I_C = 2.92$. Thus, judging by the colors, \#3.038 is
placed about half way in the effective temperature between YY~Gem and
CM~Dra. 

An independent evaluation of the relation of \#3.038 to YY~Gem and
CM~Dra can be obtained from mean densities of the stars.  The published
masses and radii for YY~Gem and CM~Dra permit direct
determination of the component densities in both systems. They are: 
$3.0 \pm 0.4$ and $4.1 \pm 0.6$ g/cm$^3$ for YY~Gem 
and $21 \pm 3$ and $22 \pm 3$ g/cm$^3$ for CM~Dra. 
For \#3.038 we can use the Kepler's law,
as re-written by Mochnacki \markcite{mo81}(1981): $\rho_1 =
0.079/(V_1 (1+q) P^2)$ and $\rho_2 = 0.079 q/(V_2 (1+q) P^2)$
g/cm$^3$, with the period $P$ in days. This
form emphasizes the orbital period which is known practically with no
error. The relative volumes of components, $V_1$ and $V_2$ (in orbital
units), are obviously poorly known as they contain uncertainties in
radii in third power. Inserting our results for \#3.038, 
we obtain $\rho_1 = 4.8
\pm 0.4$ and $\rho_2 = 6.0 \pm 3.4$ g/cm$^3$. These are evaluations
based on the assumption that the errors in $q$ and $R_i$ are
uncorrelated. The bootstrap sampling estimates which explicitly take
into account the correlations  are less pessimistic: $\rho_1 =
5.1^{+0.6}_{-0.7}$ and $\rho_2 = 5.9^{+1.6}_{-1.5}$ g/cm$^3$.

Instead of using the derived geometric parameters, we can note that 
the components of \#3.038 are apparently just inside their critical 
Roche lobes. Therefore, relatively 
firm {\it lower limits\/} on the densities can be estimated 
by using the volumes of these lobes, $V_1(q)$ and $V_2(q)$. 
The only source of uncertainty in these estimates
is then in the mass-ratio $q$.
We used the formulae of Eggeleton\markcite{Egg} 
(1983) to determine the Roche-lobe
volumes, $V_i(q)$. Using the iterated solution, we 
obtained $\rho_1 \ge 4.17 \pm 0.07$ 
and $\rho_2 \ge 4.60 \pm 0.05$ g/cm$^3$, whereas the
booststrap sampling experiment gave:
$\rho_1 \ge 4.09^{+0.15}_{-0.17}$ and  
$\rho_2 \ge 4.66^{+0.10}_{-0.11}$ g/cm$^3$.

The results on the density clearly show that the densities of
the components of \#3.038 are larger than those of the
YY~Gem and that the primary component in the system 
may have relatively lower density than its companion,
which which would be consistent with it being slightly more evolved of
the two. However, it is doubtful if the densities of the stars in
\#3.038 are as high as those in 
CM~Dra because the firm lower limits are probably very close to the
actual values. Thus, we have another indication that in terms of the
location on the Main Sequence, the components of \#3.038 are 
in between those for YY~Gem and CM~Dra.

\subsection{Angular momentum of \#3.038}

The short period of \#3.038 brings up the question of the angular
momentum (AM) of the system, $H$. 
Disregarding the spin angular momenta which
an contribute about 10\% to the total AM, the formula
for the orbital angular momentum is
$H = 1.2 \times 10^{52} q/(1+q)^2 (P/1\,{\rm d})^{1/3}
(M_{tot}/M_\odot)^{5/3}$ in cgs units. For the current period
of 0.2 day, and assuming $M_{tot} = 1\,M_\odot$, the
systems contains now about $H \simeq 1.8 \times 10^{51}$ cgs.
This a very small amount, obviously 
due to the small masses and the short period. The question
is: How much AM has been lost by the system in its
evolution? The answer depends on when, at which orbital period, the
components became inter-locked in the spin-orbit synchronism. If we
assume that this happened at about $P \simeq 2$ days, which seems a
plausible number, i.e. when $H \simeq
3.9 \times 10^{51}$ cgs, and simply subtract the two values,
then the answer is that \#3.038 has lost as much
AM in evolving from 2 days to 0.2 day 
as it contains now. In order to evaluate the AM loss rate, we should know
how long has this process has taken. For this
we have no answer as we have no idea about the age and
evolutionary state of the system. However, since the AM loss must have
been appreciable, studies specifically
addressing the questions of age and evolution will definitely 
shed light on the totally unexplored matter of the AM loss for
rapidly-rotating M-dwarfs. 

The angular momentum loss (AML) is normally thought to be
related to the overall level of
magnetic activity of stars, but the relation is still poorly
known. It appears that both, the activity and the AML tend to be
``saturated'' at high rotation rates (Vilhu\markcite{OV} 1987). 
St\c epie\'n \markcite{ste95}(1995) presented a formula for the AML
in solar-type stars which explicitely relates it to the X-ray activity. 

Results of further studies of \#3.038 may have important ramifications 
ranging from the AM evolution soon
after reaching the Main Sequence (Hartmann \& Noyes
\markcite{hn87}1987, Collier Cameron \& Jianke\markcite{CJ94} 1994,
Stauffer \markcite{stau94} 1994, Soderblom\markcite{sod96} 1996)
to the famous period-gap in cataclysmic variable stars (Rappaport et
al.\ \markcite{rap83} 1983, Spruit \& Ritter\markcite{sr83} 1983, King
\markcite{king} 1994). In the case of earlier, solar type, components
the secular evolution of the orbital period, as produced by AML and
AM transfer  by spin--orbit synchronization has been studied by several
authors  (Vilhu \markcite{OV82} 1982, Maceroni and Van't Veer
\markcite{MV91} 1991, Maceroni\markcite{CM} 1992).

At present, we have no measurements of activity of \#3.038. A deep,
pointed observation with the ROSAT satellite (Guinan, private
communication\footnote{We are indebted to Dr.\ Ed Guinan for analysis
of the Rosat data.}), 
shows no X-ray emission at the place of the star with
the upper limit of about 0.003 counts/sec, which translates into the
upper limit to the X-ray flux: $f_X < 1.8 \times 10^{-13}$
erg/cm$^2$/s. 
This upper limit is unfortunately about ten times higher
than the expected level of X-ray emission from \#3.038 assuming
the 0.1\% bolometric--to--X-rays conversion efficiency which maybe
typical for most active late-type dwarfs (Fig.~17 in
Stauffer et al.\markcite{stau94} 1994):  For $I_{max} = 15.83$,
assuming the bolometric correction $BC_I \simeq 0.5$ for $V-I \simeq
2.3$ (Bessel\markcite{Be3} 1991), we obtain $m_{bol} \simeq 15.3$ and
the bolometric observed flux $f_{bol} \simeq 1.9 \times 10^{-11}$
erg/cm$^2$/s. Therefore, the expected X-ray flux, using the Stauffer
et al.\ scaling, would be about $1.9 \times 10^{-14}$ erg/cm$^2$/s. 
Additionaly, this estimate does not include the
neutral-hydrogen absorption which maybe large in this direction, at
the distance of about 400 pc. It is known (Pye et
al.\markcite{pye94} 1994), however,
that binary systems -- even wide ones -- can have
considerably elevated X-ray activity, so that the matter is definitely
not closed. 

\section{CONCLUSIONS}

We have presented analysis of the $I$-band light curve  and of
the photometric colors 
for the closest known pair of M-type dwarfs orbiting each
other with the period of less than 0.2 day. The components 
of \#3.038 appear to
have properties intermediate between those of YY~Gem and CM~Dra;
what sets them apart from these well-studied stars, is their strong
distortion. The photometric solution indicates that the more massive
component is very close to or at it critical Roche lobe. The present 
solution is however of moderate quality as the crucial parameter of
the mass-ratio is poorly constrained by the photometric data. A
spectroscopic radial-velocity study of this important system is very
much needed. 

Tightness of the orbit and the very late spectral types of the
components invoke a series of questions related to the angular
momentum loss in very close M-type dwarfs and to non-existence of
contact binaries with spectral types later than middle-K.
 Why we do not see more systems like \#3.038? Is the angular momentum
loss too strong or too weak to produce observable quantities of such
very close systems? What is the fate of the 
system with the continuing angular momentum loss? Answers to these
questions require further observational data for this interesting
system. 

\acknowledgements

We would like to thank Drs.\ Conard Dahn, Ed Guinan, 
Kazik St\c epie\'n and Brian
Beatty for useful comments, suggestions and information. The OGLE team
is thanked for the open access to their database.

The research grant from the Natural Sciences and Engineering Council
of Canada to SR is acknowledged with gratitude.

\newpage

\begin{deluxetable}{lcc}
\tablecaption{Light curve solution of BW3\_V38 \label{tab1}}
\tablewidth{0pt}
\tablehead{
\colhead{Parameter} & \colhead{Solution} & \colhead{Bootstrap}\\
\colhead{} & \colhead{mean \& std err} & \colhead{median \& 1-sigma}
}
\startdata
$i$ (degr)       & $85.7 \pm 1.2$  & $85.8 \pm 2.2$ \nl
$q$              & $0.77 \pm 0.09$ & $0.70 \pm 0.18$ \nl
$\Omega_1$       & $3.48 \pm 0.13$ & $3.40 \pm 0.32$ \nl
$\Omega_2$       & $3.55 \pm 0.25$ & $3.36 \pm 0.48$ \nl
$T_2$ ($T_1 = 3500$) & $3459 \pm 15$ & $3456 \pm 28$   \nl
$L_1/(L_1+L_2)$  & $0.597 \pm 0.036$ & $0.597 \pm 0.065$\nl
$\Delta \phi$ (phase units)& $+0.0020 \pm 0.0006$ & $+0.0018 \pm 0.0012$\nl
\tableline
$r_1$ (side)    & $0.380 \pm 0.011$ & $0.382^{+0.026}_{-0.037}$ \nl
$r_2$ (side)    & $0.322 \pm 0.050$ & $0.323^{+0.029}_{-0.039}$ \nl
$\rho_1$ (g/cm$^3$) & $4.8 \pm 0.4$ & $5.1^{+0.6}_{-0.7}$  \nl
$\rho_2$ (g/cm$^3$) & $6.1 \pm 3.4$ & $5.9^{+1.6}_{-1.5}$ \nl
$\rho_1$ (g/cm$^3$) lim.& $>4.17 \pm 0.07$ & $>4.09^{+0.15}_{-0.17}$  \nl
$\rho_2$ (g/cm$^3$) lim.& $>4.60 \pm 0.05$ & $>4.66^{+0.10}_{-0.11}$ \nl
\enddata
\tablecomments{
$r_1$, $r_2$, $\rho_1$, $\rho_2$ have been 
derived from the solution of the first 7 parameters. See the text for
explanations of the lower limits to $\rho_1$ and $\rho_2$.}
\end{deluxetable}

\end{document}